\documentclass{aa}
\usepackage{graphicx}

\begin{document}

\title{Seismology of triple-mode classical Cepheids of the Large Magellanic Cloud}

\author{P.~Moskalik\inst{1} \and W.~A.~Dziembowski\inst{2,1}}

\institute{Copernicus Astronomical Center, ul. Bartycka 18, 00-716 Warsaw, Poland
           \and
           Warsaw University Observatory, Al. Ujazdowskie 4, 00-478 Warsaw, Poland}

\offprints{P. Moskalik,\\ \email{pam@camk.edu.pl}}

\date{Received 5 October 2004/ Accepted 23 December 2004}

\abstract{We interpret the three periods detected in OGLE \object{LMC}
Cepheids \object{SC3--360128} and \object{SC5--338399} as
corresponding to the first three overtones of radial pulsations.
This interpretation imposes stringent constraints on parameters of
the stars and on their evolutionary status, which could only be
the first crossing of the instability strip. Evolutionary models
reproducing measured periods exist only in a restricted range of
metallicities ($Z=0.004 - 0.007$). The models impose an upper
limit on the extent of overshooting from the convective core.
Absolute magnitude of each star is confined to a narrow interval.
This allows to derive a new estimate of the distance to the
\object{LMC}. We obtain $m-M$ ranging from $18\fm 34$ to $18\fm
53$, with a systematic difference between the two stars of about
$0\fm 13$. The rates of period change predicted by the models are
formally in conflict with the derived observational limits, though
the uncertainities of measured $\dot{P}$ may be underestimated. If
the discrepancy is confirmed, it would constitute a significant
challenge to the stellar evolution theory.

}

\titlerunning{Seismology of triple-mode Cepheids of \object{LMC}}
\authorrunning{Moskalik \& Dziembowski}
\maketitle

\section{Introduction}

Triple-mode radial pulsators are extremely rare. For many years
AC\thinspace Andromedae, with the fundamental and the first two
overtones simultaneously excited (Fitch \& Szeidl 1976), was the
only known object of this type. Recently, two new triple-mode
variables have been identified among short period Cepheids of the
Large Magellanic Cloud (Moskalik, Ko{\l}aczkowski \& Mizerski
2004, Moskalik \& Ko{\l}aczkowski 2005). Both stars,
\object{LMC~SC3--360128} (hereafter Star~1) and
\object{LMC~SC5--338399} (hereafter Star~2), have previously been
known to pulsate in two radial modes, namely in the first and
second overtones (Soszy\'nski et~al. 2000). The analysis of
Moskalik \& Ko{\l}aczkowski (2005) clearly shows presence of the
third independent periodicity. The period ratio of $P_3/P_2 =
0.840$, the same in both Cepheids, points towards identification
of the newly detected mode as a third radial overtone. The two
stars are the first known pulsators with three consecutive radial
overtones simultaneously excited.

Even if certain forms of pulsation are rare, it does not mean that
they are unimportant. The best example is the role of
double-mode Cepheid data in stimulating the revision of stellar
opacities. Each measured period of an identified mode yields an
accurate constraint on stellar parameters. For our objects we have
three such constraints. Another advantage is the membership of
both Cepheids in the \object{LMC}. The stars have been and will be
systematically monitored. Moreover, we have reasonably good estimate
of their absolute luminosities. This estimate is helpful but not
precise, because the distance to the \object{LMC} is still
debated. In fact, we will show that seismological analysis of the
two triple-mode Cepheids yields an independent distance
determination.

\begin{table*}
\centering
\caption[]{Triple-mode Cepheids in the \object{LMC}}
\begin{tabular} {lccccccccc}
\hline
\hline
\noalign{\smallskip}
\multicolumn{1}{l}{OGLE No.}  &
\multicolumn{1}{c}{}          &
\multicolumn{1}{c}{$P_1$}     &
\multicolumn{1}{c}{$P_2$}     &
\multicolumn{1}{c}{$P_3$}     &
\multicolumn{1}{c}{$V$}       &
\multicolumn{1}{c}{$I$}       &
\multicolumn{1}{c}{$E(B-V)$}  &
\multicolumn{1}{c}{$(V-I)_0$} &
\multicolumn{1}{c}{$W_I$}     \\

\multicolumn{1}{c}{}      &
\multicolumn{1}{c}{}      &
\multicolumn{1}{c}{[day]} &
\multicolumn{1}{c}{[day]} &
\multicolumn{1}{c}{[day]} &
\multicolumn{1}{c}{[mag]} &
\multicolumn{1}{c}{[mag]} &
\multicolumn{1}{c}{[mag]} &
\multicolumn{1}{c}{[mag]} &
\multicolumn{1}{c}{[mag]} \\
\noalign{\smallskip}
\hline
\noalign{\smallskip}
  \object{LMC SC3--360128} && 0.541279 & 0.436049 & 0.366301 & 17.53 & 16.98 & 0.134 & 0.38 & 16.13 \\
  \object{LMC SC5--338399} && 0.579510 & 0.466624 & 0.392116 & 17.37 & 16.87 & 0.133 & 0.33 & 16.09 \\
\noalign{\smallskip}
\hline
\end{tabular}
\label{data}
\end{table*}

Our aim is to extract maximum information on the two stars from
all the available data. After summarizing observational
information for the two objects (Section~2), we focus on the
inference from the values of measured periods. Our analysis in
Section~3 is similar to that of Kov\'acs \& Buchler (1994) for
AC\thinspace And. We conclude, that both triple-mode Cepheids must
be in the evolutionary phase of the Hertzsprung gap crossing and
we confront seismologically derived stellar parameters with
predictions of the appropriate evolutionary models (Section~4).
This further constrains the ranges of admissible stellar
parameters. Section~5 is devoted to the \object{LMC} distance
estimation implied by our models. In Section~6 we compare
theoretically predicted rates of period change with the limits
inferred from observations. Finally, in Section~7 we summarize our
conclusions and discuss future observations.

\section{Data}

Main observational properties of both triple-mode Cepheids are
summarized in Table~\ref{data}. Pulsation periods are determined
from OGLE $I$-band photometry reduced with the DIA algorithm
(\.Zebru\'n et~al. 2001). All periods are accurate to better than
$7\times 10^{-6}$\thinspace day. Their errors are not quoted here;
we believe they are smaller than uncertainities of the model
periods and, therefore, they are inconsequential for our analysis.
The mean $V$-band and $I$-band magnitudes are taken from
Soszy\'nski et~al. (2000). The average $E(B-V)$ reddening for the
field of each Cepheid (Udalski et~al. 1999) is used to obtain
deredenned $(V-I)_0$ colours. Standard extinction coefficients are
adopted: $A_V = 3.24\, E(B-V), A_I = 1.96\, E(B-V)$. In the last
column of the table we give Wessenheit index, $W_I = I -
1.55\times (V-I)$, which is an extinction insensitive brightness
indicator (Madore \& Freedman 1991).

\begin{figure*}
\resizebox{\hsize}{!}{\includegraphics{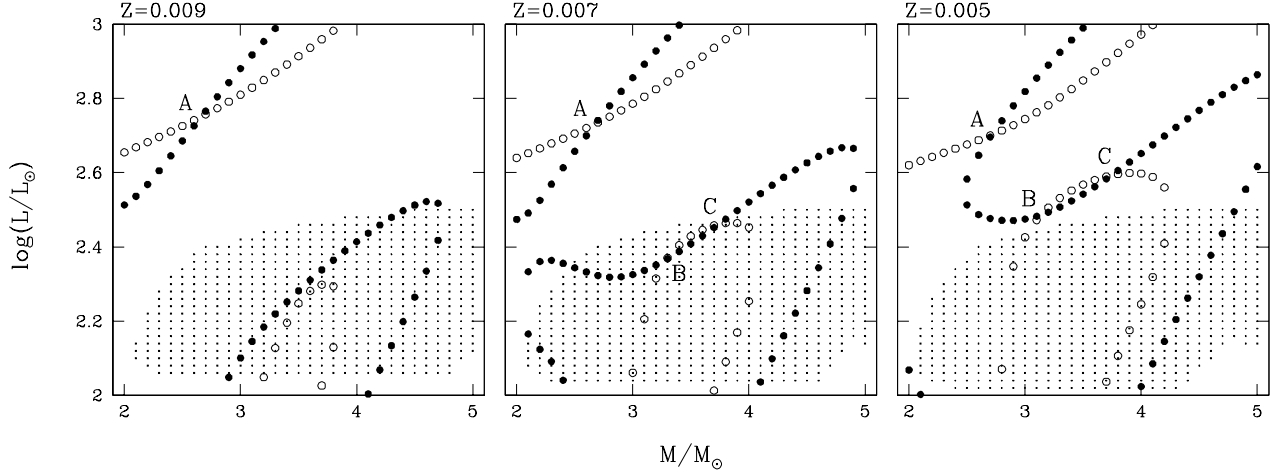}}
\caption{Simultaneous fit to three pulsation periods observed in
Star~2. Three panels display models calculated for $\alpha_{\rm
con} = 0$, $X = 0.70$ and metallicities of $Z = 0.009$, $Z=0.007$
and $Z=0.005$. Filled and open circles denote lines of $\Delta P_2
= 0$ and $\Delta P_3 = 0$, respectively. Different triple-mode
solutions are marked $A$, $B$ and $C$. Dotted area is the domain
where all three modes are linearly unstable.}
\label{one}
\end{figure*}

\begin{table*}
\centering
\caption[]{Seismic envelope models for Star~1}
\begin{tabular} {cccccccccc }
\hline
\hline
\noalign{\smallskip}
\multicolumn{1}{c}{$\alpha_{\rm con}$}  &
\multicolumn{1}{c}{$X$}                 &
\multicolumn{1}{c}{$Z$}                 &
\multicolumn{1}{c}{}                    &
\multicolumn{1}{c}{$M/M_{\odot}$}       &
\multicolumn{1}{c}{$\log(L/L_{\odot})$} &
\multicolumn{1}{c}{$\log T_{\rm eff}$}  &
\multicolumn{1}{c}{$(V-I)$}             &
\multicolumn{1}{c}{$m-M$}               \\

\multicolumn{1}{c}{}      &
\multicolumn{1}{c}{}      &
\multicolumn{1}{c}{}      &
\multicolumn{1}{c}{}      &
\multicolumn{1}{c}{}      &
\multicolumn{1}{c}{}      &
\multicolumn{1}{c}{[K]}   &
\multicolumn{1}{c}{[mag]} &
\multicolumn{1}{c}{[mag]} \\
\noalign{\smallskip}
\hline
\noalign{\smallskip}
  0 & 0.74 & 0.006 &&  3.29  & 2.311 &  3.814  & 0.50 & 18.43 \\
    &      &       &&  3.63  & 2.403 &  3.829  & 0.43 & 18.52 \\
\noalign{\smallskip}
  0 & 0.74 & 0.005 &&  3.15  & 2.358 &  3.828  & 0.44 & 18.42 \\
    &      &       && (3.67~ & 2.487 & ~3.849) &      &       \\
\noalign{\smallskip}
\hline
\noalign{\smallskip}
  1 & 0.74 & 0.007 &&  3.48  & 2.299 &  3.807  & 0.52 & 18.47 \\
    &      &       &&  3.67  & 2.359 &  3.818  & 0.48 & 18.52 \\
\noalign{\smallskip}
  1 & 0.74 & 0.006 &&  3.21  & 2.316 &  3.817  & 0.48 & 18.42 \\
    &      &       &&  3.68  & 2.423 &  3.834  & 0.41 & 18.53 \\
\noalign{\smallskip}
  1 & 0.74 & 0.005 &&  3.14  & 2.376 &  3.833  & 0.42 & 18.41 \\
    &      &       &&  3.59  & 2.461 &  3.845  & 0.37 & 18.51 \\
\noalign{\smallskip}
  1 & 0.74 & 0.004 &&  3.14  & 2.433 &  3.847  & 0.36 & 18.42 \\
    &      &       && (3.60~ & 2.547 & ~3.865) &      &       \\
\noalign{\smallskip}
\hline
\noalign{\smallskip}
  0 & 0.70 & 0.008 &&  3.28  & 2.195 &  3.787  & 0.60 & 18.39 \\
    &      &       &&  3.48  & 2.263 &  3.800  & 0.55 & 18.45 \\
\noalign{\smallskip}
  0 & 0.70 & 0.007 &&  3.10  & 2.215 &  3.795  & 0.57 & 18.37 \\
    &      &       &&  3.63  & 2.366 &  3.822  & 0.46 & 18.50 \\
\noalign{\smallskip}
  0 & 0.70 & 0.006 &&  2.98  & 2.279 &  3.813  & 0.50 & 18.36 \\
    &      &       &&  3.66  & 2.437 &  3.839  & 0.39 & 18.52 \\
\noalign{\smallskip}
  0 & 0.70 & 0.005 &&  2.88  & 2.346 &  3.832  & 0.42 & 18.35 \\
    &      &       && (3.66~ & 2.506 & ~3.856) &      &       \\
\noalign{\smallskip}
\hline
\noalign{\smallskip}
  1 & 0.70 & 0.007 &&  3.21  & 2.240 &  3.799  & 0.56 & 18.40 \\
    &      &       &&  3.67  & 2.388 &  3.827  & 0.44 & 18.51 \\
\noalign{\smallskip}
  1 & 0.70 & 0.006 &&  2.96  & 2.300 &  3.819  & 0.47 & 18.36 \\
    &      &       &&  3.65  & 2.440 &  3.840  & 0.39 & 18.51 \\
\noalign{\smallskip}
  1 & 0.70 & 0.005 &&  2.90  & 2.362 &  3.836  & 0.40 & 18.35 \\
    &      &       && (3.61~ & 2.487 & ~3.852) &      &       \\
\noalign{\smallskip}
  1 & 0.70 & 0.004 &&  2.86  & 2.410 &  3.848  & 0.35 & 18.35 \\
    &      &       && (3.63~ & 2.575 & ~3.873) &      &       \\
\noalign{\smallskip}
\hline
\end{tabular}
\label{seis1}
\end{table*}

\section{Constraints on stellar parameters from pulsation periods}

The basic assumption behind our analysis is that the three periods
detected in the triple-mode Cepheids correspond to the first three
consecutive overtones of radial pulsations. The constraints on the
stellar parmeters are derived by equating the three measured
periods $P_{n,o}$ with periods calculated in plausible stellar
models, $P_{n,c}$. In this section we use only envelope models,
extending from the photosphere down to the temperature of
$\log T = 7.2$. This is the standard approach in studies of
radial pulsations in giants. We checked that the calculated
periods are sufficently insensitive to the location of the inner
boundary of the envelope. The model periods depend on six
parametrs. These are: the mass, $M$, luminosity, $L$, effective
temperature, $T_{\rm eff}$, the fractional abundances of hydrogen,
$X$, and of elements heavier than helium, $Z$, and the mixing
length parameter, $\alpha_{\rm con}$. The pulsation data yield for
each star three equations of the form

\vskip -5pt

$$P_{n,c}(M,L,T_{\rm eff},X,Z,\alpha_{\rm con})=P_{n,o}$$

\noindent and this is not enough to determine the values of all
the unknown parameters. In principle, we could use mean magnitudes
and colours of the stars, but these are uncertain. Therefore, we
decided to find solutions for $M$, $L$ and $T_{\rm eff}$ for
selected $X$, $Z$, and $\alpha_{\rm con}$ in admissible ranges,
relying only on the measured periods and requiring that all three
modes are linearly unstable at specified parameters. Comparison
with the observed mean colours and magnitudes is done {\it a
posteriori}.

Our static stellar models, both envelope models used here and the
evolutionary models used in Section~4.2, are calculated with a
modernized version of the Paczy\'nski (1970) codes. The main
improvements consist in the implementation of the OPAL
equation-of-state and opacity data (Rogers, Swenson \& Iglesias
1996; Iglesias \& Rogers 1996). Pulsation periods are calculated
with a modified version of the Dziembowski (1977) code.

There are uncertainties in the calculated periods. We rely on the
linear nonadiabatic approximation. We believe this is the most
severe simplification. Comparison of periods derived from linear
and from nonlinear codes has been published only for monoperiodic
RR~Lyr models (e.g. Kov\'acs \& Buchler 1988) and long period
Cepheid models (e.g. Moskalik \& Buchler 1991; Fokin 1994). It is
not clear how to scale these results to higher overtones in
multiperiodic $\sim 3M_{\odot}$ Cepheids.

Errors may result from using the mixing length theory to calculate
convective flux in our static models and ignoring its Lagrangian
perturbation in pulsating models. We believe that this is less
essential and that our calculations made with $\alpha_{\rm con}=0$
and $\alpha_{\rm con}=1$ reflect well the uncertainty. The use of
the Edddington approximation in the optically shallow layers is
also an inadequacy but of secondary importance for the present
application.

Simultaneous fit to three radial mode periods can be achieved for
several different sets of parameters ({\it e.g.} AC\thinspace And;
Kov\'acs \& Buchler 1994). In order to assure that we find {\it
all} the solutions we apply the following procedure:

\begin{enumerate}
\item first, we select the chemical composition ($X$, $Z$) and the
      mixing length $\alpha_{\rm con}$. Keeping these three parameters
      fixed, we calculate a grid of stellar models covering the mass
      range of $2 < M/M_{\odot} < 5$ and luminosity range of
      $2 < \log(L/L_{\odot}) < 3$.

\item at every point of the ($M$, $L$) grid we find a model which
      fits the observed period of the first overtone, $P_{1,o}$. This
      can be easily achieved by adjusting the last remaining parameter
      of the model, namely its effective temperature. The solution for
      $T_{\rm eff}$ is unique. The model fitting $P_1$ does not necesarily
      reproduce the other two periods. We calculate period mismatches
      $\Delta P_2=P_{2,c} - P_{2,o}$ and $\Delta P_3=P_{3,c} - P_{3,o}$.

\item we find on the ($M$, $L$) plane the lines of $\Delta P_2 = 0$.
      Models located on these lines reproduce {\it both} $P_1$ and $P_2$.
      Similarly, we find the lines of $\Delta P_3 = 0$. The values of
      ($M$, $L$) at which the two types of lines cross each other define
      the models which simultaneously reproduce {\it all three periods}.
\end{enumerate}

\begin{table*}
\centering
\caption[]{Seismic envelope models for Star~2}
\begin{tabular} {cccccccccc}
\hline
\hline
\noalign{\smallskip}
\multicolumn{1}{c}{$\alpha_{\rm con}$}  &
\multicolumn{1}{c}{$X$}                 &
\multicolumn{1}{c}{$Z$}                 &
\multicolumn{1}{c}{}                    &
\multicolumn{1}{c}{$M/M_{\odot}$}       &
\multicolumn{1}{c}{$\log(L/L_{\odot})$} &
\multicolumn{1}{c}{$\log T_{\rm eff}$}  &
\multicolumn{1}{c}{$(V-I)$}             &
\multicolumn{1}{c}{$m-M$}               \\

\multicolumn{1}{c}{}      &
\multicolumn{1}{c}{}      &
\multicolumn{1}{c}{}      &
\multicolumn{1}{c}{}      &
\multicolumn{1}{c}{}      &
\multicolumn{1}{c}{}      &
\multicolumn{1}{c}{[K]}   &
\multicolumn{1}{c}{[mag]} &
\multicolumn{1}{c}{[mag]} \\
\noalign{\smallskip}
\hline
\noalign{\smallskip}
  0 & 0.70 & 0.008 &&  3.45  & 2.333 &  3.807  & 0.52 & 18.52 \\
    &      &       &&  3.65  & 2.382 &  3.815  & 0.49 & 18.57 \\
\noalign{\smallskip}
  0 & 0.70 & 0.007 &&  3.29  & 2.367 &  3.818  & 0.48 & 18.50 \\
    &      &       &&  3.74  & 2.461 &  3.833  & 0.42 & 18.60 \\
\noalign{\smallskip}
  0 & 0.70 & 0.006 &&  3.21  & 2.424 &  3.834  & 0.41 & 18.49 \\
    &      &       && (3.74~ & 2.525 & ~3.848) &      &       \\
\noalign{\smallskip}
\hline
\noalign{\smallskip}
  1 & 0.70 & 0.008 &&  3.41  & 2.337 &  3.809  & 0.51 & 18.51 \\
    &      &       &&  3.74  & 2.416 &  3.823  & 0.46 & 18.58 \\
\noalign{\smallskip}
  1 & 0.70 & 0.007 &&  3.27  & 2.385 &  3.824  & 0.45 & 18.49 \\
    &      &       &&  3.75  & 2.469 &  3.836  & 0.40 & 18.59 \\
\noalign{\smallskip}
  1 & 0.70 & 0.006 &&  3.25  & 2.438 &  3.837  & 0.40 & 18.50 \\
    &      &       &&  3.67  & 2.504 &  3.845  & 0.36 & 18.59 \\
\noalign{\smallskip}
  1 & 0.70 & 0.005 &&  3.26  & 2.489 &  3.849  & 0.35 & 18.51 \\
    &      &       && (3.64~ & 2.563 & ~3.860) &      &       \\
\noalign{\smallskip}
\hline
\noalign{\smallskip}
  0 & 0.72 & 0.007 &&  3.52  & 2.397 &  3.822  & 0.46 & 18.54 \\
    &      &       &&  3.66  & 2.430 &  3.827  & 0.44 & 18.58 \\
\noalign{\smallskip}
  0 & 0.72 & 0.006 &&  3.38  & 2.438 &  3.834  & 0.41 & 18.53 \\
    &      &       && (3.72~ & 2.508 & ~3.845) &      &       \\
\noalign{\smallskip}
\hline
\noalign{\smallskip}
  1 & 0.72 & 0.007 &&  3.42  & 2.399 &  3.824  & 0.45 & 18.53 \\
    &      &       &&  3.73  & 2.456 &  3.832  & 0.42 & 18.60 \\
\noalign{\smallskip}
  1 & 0.72 & 0.006 &&  3.45  & 2.458 &  3.838  & 0.39 & 18.54 \\
    &      &       &&  3.59  & 2.482 &  3.841  & 0.38 & 18.57 \\
\noalign{\smallskip}
\hline
\end{tabular}
\label{seis2}
\end{table*}

In Fig.\,\ref{one} we display the results of our procedure applied
to Star~2, for $\alpha_{\rm con} = 0$, $X = 0.70$ and three
different metallicity values of $Z = 0.009$, 0.007 and 0.005. The
filled and open circles define the lines of $\Delta P_2 = 0$ and
$\Delta P_3 = 0$, respectively. The dotted area in each plot
corresponds to models with all three modes linearly unstable. We
recall here, that in the absence of resonances (as in our case),
the linear excitation of all modes is the necessary condition for
the triple-mode pulsations to exist. Among all models matching the
observed periods, only those located in the dotted area correspond
to acceptable solutions.

As we see from Fig.\,\ref{one}, the problem of matching
periods of a triple-mode Cepheid can have up to three different
solutions. We call them $A$, $B$ and $C$. Solution $A$ exists for
all choices of $X$, $Z$ and $\alpha_{\rm con}$, but all three
modes are always linearly stable. We will not discuss this
solution any further. Solutions $B$ and $C$ can exist only for
metallicities below a certain critical value. This is clearly seen
in Fig.\,\ref{one}: for $Z=0.009$ the lines $\Delta P_2 = 0$ and
$\Delta P_3 = 0$ are close to each other, but do not cross. As the
value of $Z$ is lowered, the two lines come closer and finally
cross each other, generating solutions $B$ and $C$. For the case
displayed in Fig.\,\ref{one}, this happens for $Z\sim 0.0082$.
When $Z$ is lowered further, first solution $C$ and then solution
$B$ migrates out of the doted area. This sets a {\it lower limit}
for the metallicity. Below this limit solutions $B$ and $C$ do
exist, but do not correspond to triple-mode pulsations, because at
least one of the modes is linearly stable. For the case of
Fig.\,\ref{one} this lower limit is at $Z\sim 0.0056$.

We have performed search for solutions for both triple-mode
Cepheids, exploring the range of metallicities from $Z=0.004$ to
0.010 (in some cases to 0.020) and of hydrogen abundances from
$X=0.70$ to 0.74. For each chemical composition we have calculated
models with $\alpha_{\rm con} = 0$ and with $\alpha_{\rm con} =1$.
The results of this survey are summarized in Table~\ref{seis1} for
Star~1 and in Table~\ref{seis2} for Star~2. We list only those
combinations of $X$, $Z$ and $\alpha_{\rm con}$ for which at least
one acceptable solution exists. For low metallicities, solution
$C$ may already fall outside the "all-modes-unstable" domain, in
such a case it is given in parenthesis. For all acceptable
solutions we provide the values of $(V-I)$ colour and the distance
modulus to the star, $m-M$. The latter is obtained by comparing
Wessenheit index, $W_I$, derived from the models with the mean
observed values, as given in  Table~\ref{data}. The conversion
from model parameters to $V$-band and $I$-band magnitudes is done
with tabular data based on Kurucz's (1998) stellar atmosphere
models.

The triple-mode pulsations strongly constrain parameters of the
studied Cepheids. We find, that metallicities of both stars have
to be in the range of $Z=0.0035-0.008$, with limiting values
somewhat dependent on $X$ and assumed $\alpha_{\rm con}$. For
Star~1, fitting the observed periods is possible for all values of
$X$, but for Star~2 hydrogen abundance of $X \leq 0.72$ is
required. The inferred stellar masses are limited to ranges of
$2.8 < M/M_{\odot} < 3.7$ for Star~1 and $3.2 < M/M_{\odot} < 3.8$
for Star~2. The respective distance moduli are in the ranges of
$18\fm 35-18\fm 53$ and $18\fm 49-18\fm 60$.

In the next section we will show that the constraints imposed by
triple-mode pulsations can be considerably tightened by combining
the seismic envelope models with the appropriate evolutionary
tracks.

\section{Evolutionary models}

\subsection{Evolutionary status of the stars}

The very short periods and low luminosities of the triple-mode
Cepheids cannot be explained with the current models of stars in
the phase of central helium burning. Whether or not a star enters
the instability strip during this evolutionary phase depends on
the extent of {\it the blue loop}. The latter is mainly determined
by the mass and the metallicity of the star. Alibert et~al. (1999)
find that at $Z=0.004$ their models with masses of $3.0-3.5
M_\odot$ do enter the instability strip. However, the minimum
value of $P_1$, which occurs for $M=3M_\odot$ is about $0\fd 9$.
At $Z=0.01$ they find models with blue loop extending to the
instability strip only at $M\ge 4M_\odot$. Such models have,
naturally, much higher luminosities and pulsation periods than our
stars. Alibert et~al. (1999) assumed negligible overshooting in
their calculations.

The effect of finite overshooting was taken into account in an
extensive set of models of Girardi et~al. (2000). In this case,
the parameters we have derived for the triple-mode Cepheids are
even more difficult to reconcile with the models in the helium
burning phase. At $Z=0.004$ and $M=3M_\odot$ the loop does not
extend to the instability strip. At $M=3.5M_\odot$ the loop ends
within the strip, but then the minimum value of $P_1$ is about
$1\fd 15$. Again, increasing $Z$ only worsens the discrepancy.

Determination of the extent of the blue loop is a subtle problem.
Here we find the largest discrepancies between various authors. A
more robust result is the difference of $\Delta\log L\sim 0.2$
between the first and the second crossing of the instability strip
by stars in the $3-4 M_\odot$ mass range. The $\log L$ values
we find for the triple-mode Cepheids indicate that these two
objects are caught during the short phase of the first crossing of
the strip.

\subsection{Models in the first crossing of the instability strip}

We have computed a number of evolutionary tracks covering the
Hertzsprung gap crossing phase. From each track, we may determine
$L$ in the moment when the first overtone period is equal to
$P_{1,o}$. The value of $L$ is virtually independent of
$\alpha_{\rm con}$, but depends strongly on $\alpha_{\rm ov}$,
which is the ratio of the overshooting distance from the
convective core to the pressure distance scale calculated at the
edge of the core. From our model calculations we find

$$\left({\partial\log L\over\partial\alpha_{\rm ov}}\right)_{M,X,Z,P_1} \approx0.6$$

\vskip 8pt

\noindent At specified values of $M$, $X$, $Z$ and $P_1$, the
minimum value of $L$ corresponds to $\alpha_{\rm ov}=0$

\begin{table*}
\centering
\caption[]{Triple-mode Cepheid models consistent with no overshooting evolutionary tracks}
\begin{tabular} {cccccccccc}
\hline
\hline
\noalign{\smallskip}
\multicolumn{1}{c}{$X$}                   &
\multicolumn{1}{c}{$\alpha_{\rm con}$}    &
\multicolumn{1}{c}{}                      &
\multicolumn{1}{c}{$Z$}                   &
\multicolumn{1}{c}{$M/M_{\odot}$}         &
\multicolumn{1}{c}{$\log\,(L/L_{\odot})$} &
\multicolumn{1}{c}{$\log T_{\rm eff}$}    &
\multicolumn{1}{c}{$(V-I)$}               &
\multicolumn{1}{c}{$m-M$}                 &
\multicolumn{1}{c}{$\dot{P}/P$}           \\

\multicolumn{1}{c}{}             &
\multicolumn{1}{c}{}             &
\multicolumn{1}{c}{}             &
\multicolumn{1}{c}{}             &
\multicolumn{1}{c}{}             &
\multicolumn{1}{c}{}             &
\multicolumn{1}{c}{[K]}          &
\multicolumn{1}{c}{[mag]}        &
\multicolumn{1}{c}{[mag]}        &
\multicolumn{1}{c}{[Myr$^{-1}$]} \\
\noalign{\smallskip}
\hline
\noalign{\smallskip}
\multicolumn{10}{c}{Star~1} \\
\noalign{\smallskip}
\hline
\noalign{\smallskip}
  0.70 & 0 & \ & 0.0061 & 2.99 & 2.273 & 3.811 & 0.51 & 18.36 & 1.51 \\
       & 1 & \ & 0.0063 & 3.02 & 2.280 & 3.813 & 0.50 & 18.37 & 1.53 \\
\noalign{\smallskip}
  0.74 & 0 & \ & 0.0057 & 3.25 & 2.323 & 3.818 & 0.48 & 18.43 & 1.65 \\
       & 1 & \ & 0.0062 & 3.26 & 2.308 & 3.814 & 0.50 & 18.43 & 1.61 \\
\noalign{\smallskip}
\hline
\noalign{\smallskip}
\multicolumn{10}{c}{Star~2} \\
\noalign{\smallskip}
\hline
\noalign{\smallskip}
  0.70 & 0 & \ & 0.0065 & 3.24 & 2.394 & 3.826 & 0.44 & 18.49 & 2.15 \\
       & 1 & \ & 0.0069 & 3.26 & 2.392 & 3.826 & 0.44 & 18.49 & 2.18 \\
\noalign{\smallskip}
  0.72 & 0 & \ & 0.0061 & 3.39 & 2.432 & 3.832 & 0.42 & 18.53 & 2.11 \\
       & 1 & \ & ---    & ---  & ---   & ---   & ---  & ---   & ---  \\
\noalign{\smallskip}
\hline
\end{tabular}
\label{evol0}
\end{table*}

\begin{figure}
\resizebox{\hsize}{!}{\includegraphics{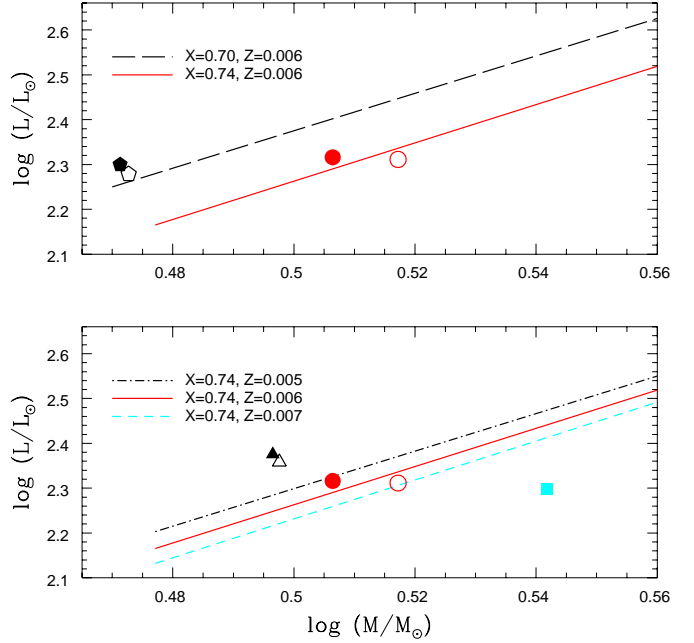}}
\caption{Evolutionary $M-L$ relation for models with $P_1 = 0\fd
541$. Lines are derived for stellar models undergoing first
crossing of the instability strip. Models are calculated assuming
no overshooting from the convective core ($\alpha_{\rm ov}=0$)
during the main sequence evolution. Symbols represent seismic
solutions for Star~1, derived with (non-evolutionary) envelope
models (see Table~\ref{seis1}). Filled and open symbols correspond
to $\alpha_{\rm con}=1$ and 0, respectively. Pentagons correspond
to $X=0.70, Z=0.006$, circles to $X=0.74, Z=0.006$, triangles to
$X=0.74, Z=0.005$, and square to $X=0.74, Z=0.007$ (there is no
solution for $\alpha_{\rm con}=0$ in the last case).}
\label{two}
\end{figure}

In Fig.\,\ref{two} we display the $M-L$ relations for no
overshooting evolutionary models of Star~1 ($P_1 = 0\fd 541$).
Different lines correspond to various chemical compositions. In
the same plot we also display parameters of several seismic models
of Star~1, also for different values of $X$ and $Z$ (see
Table~\ref{seis1}). We plot only models corresponding to solution
$B$, because only for this solution agreement between seismic and
evolutionary models can be reached. The lower panel of the plot
explores the effect of varying stellar metallicity. For $Z=0.005$
the seismic models do not agree with $\alpha_{\rm ov}=0$
evolutionary tracks. The agreement may be easily reached if
overshooting of $\alpha_{\rm ov}\sim 0.1$ is allowed, which shifts
the evolutionary $M-L$ relation up. For $Z=0.006$ the seismic
models and the evolutionary tracks are almost in agreement. For
$Z=0.007$ the seismic model is placed below the corresponding
evolutionary $M-L$ line and no agreement between the two is
possible. Fig.\,\ref{two} illustrates the general pattern of model
behaviour: with $X$, $\alpha_{\rm con}$ and $\alpha_{\rm ov}$
fixed, the exact match between seismic models and evolutionary
tracks can be reach only at one specific metallicity value. The
matching values of $Z$ becomes lower, when the overshooting
distance $\alpha_{\rm ov}$ is increased.

The upper panel of Fig.\,\ref{two} shows that match between
seismic and evolutionary models is possible for any reasonable
value of $X$, but yielding different values of stellar mass:
$M\sim 3.2M_\odot$ for $X=0.74$ and $M\sim 3.0M_\odot$ for
$X=0.70$. Interestingly, metallicities of matching models are
almost the same independently of $X$.

\begin{table*}
\centering
\caption[]{Triple-mode Cepheid models for maximum overshooting alowed by seismic constraints}
\begin{tabular} {ccccccccccc}
\hline
\hline
\noalign{\smallskip}
\multicolumn{1}{c}{$X$}                   &
\multicolumn{1}{c}{$\alpha_{\rm con}$}    &
\multicolumn{1}{c}{}                      &
\multicolumn{1}{c}{$Z$}                   &
\multicolumn{1}{c}{$\alpha_{\rm ov}$}     &
\multicolumn{1}{c}{$M/M_{\odot}$}         &
\multicolumn{1}{c}{$\log\,(L/L_{\odot})$} &
\multicolumn{1}{c}{$\log T_{\rm eff}$}    &
\multicolumn{1}{c}{$(V-I)$}               &
\multicolumn{1}{c}{$m-M$}                 &
\multicolumn{1}{c}{$\dot{P}/P$}           \\

\multicolumn{1}{c}{}             &
\multicolumn{1}{c}{}             &
\multicolumn{1}{c}{}             &
\multicolumn{1}{c}{}             &
\multicolumn{1}{c}{}             &
\multicolumn{1}{c}{}             &
\multicolumn{1}{c}{}             &
\multicolumn{1}{c}{[K]}          &
\multicolumn{1}{c}{[mag]}        &
\multicolumn{1}{c}{[mag]}        &
\multicolumn{1}{c}{[Myr$^{-1}$]} \\
\noalign{\smallskip}
\hline
\noalign{\smallskip}
\multicolumn{11}{c}{Star~1} \\
\noalign{\smallskip}
\hline
\noalign{\smallskip}
  0.70 & 0 & \ & 0.0043 & 0.33 & 2.81 & 2.393 & 3.845 & 0.36 & 18.34 & 1.69 \\
       & 1 & \ & 0.0036 & 0.33 & 2.84 & 2.428 & 3.853 & 0.33 & 18.34 & 1.81 \\
\noalign{\smallskip}
  0.74 & 0 & \ & 0.0045 & 0.22 & 3.07 & 2.392 & 3.838 & 0.40 & 18.40 & 1.61 \\
       & 1 & \ & 0.0038 & 0.19 & 3.14 & 2.443 & 3.849 & 0.35 & 18.42 & 1.75 \\
\noalign{\smallskip}
\hline
\noalign{\smallskip}
\multicolumn{11}{c}{Star~2} \\
\noalign{\smallskip}
\hline
\noalign{\smallskip}
  0.70 & 0 & \ & 0.0056 & 0.12 & 3.18 & 2.450 & 3.841 & 0.38 & 18.49 & 2.07 \\
       & 1 & \ & 0.0049 & 0.07 & 3.26 & 2.493 & 3.850 & 0.34 & 18.51 & 2.26 \\
\noalign{\smallskip}
  0.72 & 0 & \ & 0.0058 & 0.05 & 3.35 & 2.452 & 3.838 & 0.39 & 18.52 & 2.12 \\
       & 1 & \ & ---    & ---  & ---  & ---   & ---   & ---  & ---   & ---  \\
\noalign{\smallskip}
\hline
\end{tabular}
\label{evolmax}
\end{table*}

In Table~\ref{evol0} we list parameters of the two stars resulting
from matching seismic models to the evolutionary tracks calculated
for $\alpha_{\rm ov}=0$. The values obtained with two choices of
$\alpha_{\rm con}$ give some assessment of uncertainity due to
poor knowledge of convection. We regard the parameters derived at
higher $X$ to be more likely, because we do not expect any
significant He enrichment for \object{LMC} stars with $M\sim
3M_\odot$. In the last two columns we provide the values of
$(V-I)$ colour and of the distance modulus to the stars, obtained
in same way as in Tables~\ref{seis1} and \ref{seis2}. Bearing in
mind uncertainty in the mean deredenned colour, which according to
Udalski (private communication) is about $0\fm 1$, we regard the
agreement in $(V-I)$ to be quite satisfactory. The agreement may
be even better, if smaller \object{LMC} redennings of Subramaniam
(2004) are used. We will discuss the inferred values of the
distance modulus later in this paper. In the last column of
Table~\ref{evol0} we list the calculated rates of period change.
These numbers will be discussed later, too.

We stress, that with exception of $\dot{P}/P$, all the values
listed in Table~\ref{evol0} are derived from the envelope models.
The requirement of matching the seismic envelope models and the
evolutionary tracks was used only to a) reject solution $C$ and b)
to find the value of $Z$ at which the exact match is achieved.

\subsection{Constraints on the extent of overshooting}

At fixed $X$ and $\alpha_{\rm con}$, there is a minimum
metallicity value allowed by seismic models. This lower limit for
$Z$ corresponds to an {\it upper limit} for the overshooting
distance, $\alpha_{\rm ov}$. In Table~\ref{evolmax} we give the
maximum overshooting distances for both triple-mode Cepheids. We
obtain a stronger contraint for Star~2, especially if the higher
value of $X$ is adopted. Despite very similar masses, we cannot
argue that the two stars should have similar $\alpha_{\rm ov}$.
They may well differ in the rotation rate and rotation induced
mixing can mimic the convective overshooting. In any case, the
limit of $\alpha_{\rm ov} \le 0.12$ obtained for Star~2 is very
interesting.

The triple-mode Cepheids can be quite important for constraining
element mixing in moderate mass stars. To demonstrate this point,
we list in Table~\ref{evolmax} also the stellar parameters of the
two Cepheids derieved at the extreme values of $\alpha_{\rm ov}$.
These can be compared with parameters derived for $\alpha_{\rm
ov}=0$ (Table~\ref{evol0}). As overshooting is increased, the
matching models become less metal abundant and bluer. Thus, the
limits on the extent of overshooting can be further improved once
we precisely measure either metallicities or the mean deredened
colours of the two stars. We assume here, that the
composition in the envelope and in the photosphere is the same,
which is valid unless the star is a fast rotator.

\section{The distance to \object{LMC}}

The values of distance moduli listed in Tables~\ref{evol0} and
\ref{evolmax} are all between $18\fm 34$ and $18\fm 53$, {\it
i.e.} well within currently considered range for the \object{LMC}.
They are indeed very close to estimates based on analysis of RRd
stars, which yields $18\fm 53$ according to Kov\'acs (2000b) and
$18\fm 42$ according to Popielski et~al. (2000). From the
double-mode Cepheid data, Kov\'acs (2000a) obtained the value of
$18\fm 54$. All these determinations rely on fitting the observed
pulsation periods, but for different objects and in different
ways.

For the same model parameters ($X$, $\alpha_{\rm con}$), there is
a systematic difference of about $0\fm 13$ between distance moduli
derived for Star~1 and Star~2. We believe, that large part of this
difference can be attributed to $W_I$ measurement errors. Indeed,
judging from faint Cepheids ($W_I > 15\fm 0$) with independent
photometry from two OGLE fields, the error of the mean Wessenheit
index is about $0\fm 06$ {\it on average}, but sometimes can
exceed $0\fm 09$. We cannot exclude, however, that some part of
the $0\fm 13$ difference might be real. As shown by Subramaniam
(2003), the $m-M$ displays considerable variation even within the
\object{LMC} bar.

We want to stress, that distances of the triple-mode Cepheids can
be estimated very accurately once photometric precision is
improved. As seen from Tables~\ref{evol0} and \ref{evolmax}, the
total range of $m-M$ allowed by seismic models is only $0\fm 09$
for Star~1 and $0\fm 04$ for Star~2. In principle, the individual
distances of the two stars can be determined with, respectively,
$\pm 2$\% and $\pm 1$\% accuracy.

\section{Period changes}

The first crossing of the instability strip is by two orders of
magnitude faster than either the second or the third crossing.
Measuring the rates of period change should, therefore, provide a
crucial test for the triple-mode Cepheid models.

The rates of period change, $\dot{P}/P$, inferred from our models
are given in the last column of Tables~\ref{evol0} and
\ref{evolmax}. They are about 2\thinspace Myr$^{-1}$, somewhat
below this value for Star~1 and somewhat above it for Star~2. The
rates are calculated for the first overtone, but they are
approximately the same for all modes. With such values of
$\dot{P}/P$, the $(O-C)$ should reach $6-10$\% of the pulsation
period in ten years.

We do not have yet ten years worth of data at our disposal, but we
will try to measure $\dot{P}$ with photometry already at hand. Our
method consists in the least square fitting of the photometric
measurements with the Fourier sum of the form

\vskip -8pt

$$I(t) =  \langle I\rangle + \sum_{i=1}^{3} {\rm A}_i \cos [(\omega_i  + {1\over 2}\dot\omega_i t)\,t + \phi_i] \ + \ \ \ \ \ \ \ \ \ \ \ \ \ \ \ $$
         $$ \ \ \ \ \ \ \ \ \ \ \ \ \ \ \ \ \ \ + \ harmonics + combination \ terms$$

\vskip +8pt

\noindent where $\omega_i = 2\pi/P_i$ are pulsation frequencies
and $\dot\omega_i = -2\pi\dot{P}_i/P_i^2$ are the rates of
frequency change (see {\it e.g.} Kepler 1993 for derivation). The
nonlinear least square fit is performed with the Marquardt
algorithm (Press et~al. 1986). In the following, we will discus
only the rates of period change for the first overtone. The other
two modes have much lower amplitudes and consequently their
$\dot{P}$ is determined with 3-10 times larger errors.

The method, when applied to DIA-reduced OGLE $I$-band photometry
yields $\dot{P}_1/P_1 = + 1.7\pm 3.7$\thinspace Myr$^{-1}$ for
Star~1 and $\dot{P}_1/P_1 = + 5.3\pm 6.4$\thinspace Myr$^{-1}$ for
Star~2. Clearly, the OGLE timebase of $\sim 1200$\thinspace day is
not sufficient to obtain significant $\dot{P}$ measurement.

A considerably longer timebase ($\sim 2700$\thinspace day) is
ofered by MACHO photometry (Allsman \& Axelrod 2001). Both
triple-modes Cepheids have been observed by MACHO Project
(\object{LMC SC3--360128}\thinspace
=\thinspace\object{MACHO\thinspace 77.8029.310}; \object{LMC
SC5--338399}\thinspace =\thinspace\object{MACHO\thinspace
80.7074.120}) and over 1500 measurements have been accumulated for
each of them. Unfortunately, MACHO data are of much lower quality.
We have decided to perform initial filtering of MACHO photometry,
by rejecting a)~all outlying measurements deviating by more than
$5\sigma$ from average brightness of the star and b)~all
measurements with photometric errors (given by MACHO) larger that
5 times the averge for the dataset. Further data clipping is done
during the fitting procedure, when all points deviating by more
than $5\sigma$ from the fit are also rejected. For the final
determination of $\dot{P}$ we use the MACHO $B$-band photometry.
The analysis yields

\vskip -5pt

$$\dot{P}_1/P_1 = - 1.66\pm 0.83\, {\rm Myr}^{-1} \ \ \ \ {\rm for \ Star \ 1}$$

\vskip -20pt

$$\dot{P}_1/P_1 = - 2.28\pm 0.94\, {\rm Myr}^{-1} \ \ \ \ {\rm for \ Star \ 2}$$

\vskip +8pt

\noindent For both Cepheids $\dot{P}/P$ is determined only with
$2\sigma$ accuracy. Therefore, we do not consider the above values
to be significant detections, but rather to provide observational
limits on $\dot{P}$. The results, nevertheless, are very
interesting, because these limits seem to be in conflict with the
model predictions. The measured $\dot{P}/P$ values differ from
theoretical ones (Tables~\ref{evol0} and \ref{evolmax}) by
$3.8\sigma$ for Star~1 and by $4.6\sigma$ for Star~2,
respectively. If real, such a discrepancy would mean serious
troubles for the evolutionary models. However, we should treat
this result with a grain of salt. The quoted errors of $\dot{P}/P$
are formal statistical errors resulting from the nonlinear least
square procedure. Such errors very often underestimate the true
uncertainities of the fited parameters. This is especially true in
the case of multiperiodic pulsations (Costa, Kepler \& Winget
1999; Costa \& Kepler 2000).

The $\dot{P}$ determinations reported in this section require
confirmation with better data. This will soon be possible when
OGLE~III photometry is released. Combined OGLE~II+III dataset will
provide the timebase similar to that of MACHO, but much higher
data quality should result in $\dot{P}$ errors about twice
smaller.

Further accumulation of data will eventually allow determinaton of
$\dot{P}$ for all three modes. This is quite essential for
understanding the true nature of the period changes. If they are
caused by stellar evolution, then all modes should yield roughly
the same values of $\dot{P}/P$. No such equality is expected if
period changes result from nonlinear mode interactions.

\section{Conclusions and discussion}

We show that pulsation periods measured in two \object{LMC}
triple-mode Cepheids are consistent with values calculated for the
first three overtones only if the envelope metal abundance is
between $Z=0.004$ and 0.007. This is somewhat lower than
$Z=0.008$, typically adopted for the \object{LMC}. In one of the
stars, fitting the periods requires the envelope hydrogen
abundance $X \la 0.72$, which is again somewhat below the commonly
accepted \object{LMC} value of $X=0.74$. The conclusions regarding
envelope chemical composition rests mainly on the measured period
ratios. For calculating the model periods, we rely on the linear
pulsation theory. We believe, that nonlinear corrections to the
periods are small, but we cannot estimate them. Therefore, it is
important that the derived metallicity limits are verified through
direct spectroscopic determination.

Effective temperatures implied by the models are consistent with
observed $(V-I)$ colours of the two stars. Unfortunately, the
accuracy of the latter is currently too low to yield usefull
constraints. Our models have luminosities confined to narrow
ranges. The inferred distance moduli are $18\fm 34 - 18\fm 43$ for
one star and $18\fm 49 - 18\fm 53$ for the other. The difference
between the two objects can be in large part attributed to
photometric errors. The derived distances to the triple-mode
Cepheids are within the range currently considered for the
\object{LMC}, but favour rather longer \object{LMC} distances in
this range.

We find that parameters inferred from the pulsation periods of the
two stars are consistent with evolutionary models undergoing the
Hertzsprung gap crossing. The consistency does not require any
overshooting from the convective core during the main sequence
evolution. Seismic models do not exclude overshooting, though, but
rather impose an upper limit on its extent. The limit is
particularly strong in case of \object{LMC~SC5--338399}. The
extent of overshooting can be further constrained once metal
abundances or, alternatively, mean deredenned colours of the stars
are presisely measured.

Both the periods and the luminosities of the triple-mode Cepheids
cannot be reconciled with evolutionary models in the phase of
central helium burning. There is nothing strange in our proposal
that these stars are undergoing Hertzsprung gap crossing. Stars
with masses of $\sim 3M_\odot$, which we find for our objects,
spend about $10^5$\thinspace years crossing the instability strip
on their way to the red giant branch. This is comparable to time
spent inside the strip by more massive stars ($M\sim 7M_\odot$)
during the central helium burning phase. The pulsation periods of
such stars are between 10 and 20\thinspace days. We observe many
\object{LMC} Cepheids in this period range.

What may be a problem for the first crossing scenario is a
possible conflict between predicted and measured rates of period
change. The currently available data are insufficient for
significant $\dot{P}$ determination. We note, however, that in
both Cepheids the most probable $\dot{P}$ values are more than
$3.8\sigma$ away from theoretical expectations. Since we are
somewhat sceptical about $\dot{P}$ measurement errors, we do not
yet consider this result to invalidate our models. But this is a
warning sign.

If the future determinations confirm the discrepancy, our models
will be in serious troubles. In our opinion, the problem would be
more severe than the one encountered by Pietrukowicz (2003) for
long period Cepheids. He has found that theoretically predicted
rates of period change are faster than the measured ones, but the
size of the effect depends on whose evolutionary tracks are used.
According to our unpublished calculations, there is no significant
conflict if one uses the tracks of Girardi et~al. (2000).

The speed of evolution during the second and third crossings of
the instability strip is affected by the element mixing in three
distinct evolutionary phases. There are uncertainties regarding
mixing efficiency in each of them. During the main sequence phase,
we do not know the extent of the convective overshooting from the
hydrogen burning core. During the red giant phase, we do not know
the efficiency of the dredge up of the helium enriched material
into the deep convective envelope. Finally, during the blue loop
phase, there is a subtle problem with treatment of the boundary of
the helium burning core. In the case of the first crossing of the
instability strip, only the uncertainty of the main sequence
overshooting affects the results, and this has been taken into
account in our calculations. We believe that our prediction for
the rates of period change is quite robust. Therefore, the
possible discrepancy between calculated and observed $\dot{P}$
would pose a very serious challenge to our understanding of
stellar evolution. Further observations of both triple-mode
Cepheids are needed to settle this important issue.

\begin{acknowledgements}
This work has been supported in part by Polish KBN grant 5~P03D~030~20.
\end{acknowledgements}

%\listofobjects

\end{document}